# Improvement of upper critical field by the compositionally-complex-alloy concept in A15 superconductor V$_3$X (X: Al, Si, Ga, Ge, Sn)

Yuki Nakahira[1], Ryosuke Kiyama[1], Aichi Yamashita[1], Hiroaki Itou[2], Akira Miura[2], Chikako Moriyoshi[3], Yosuke Goto[1], Yoshikazu Mizuguchi[1]*

1. Department of Physics, Tokyo Metropolitan University, 1-1, Minami-osawa, Hachioji 192-0397, Japan.

2. Faculty of Engineering, Hokkaido University, Kita 13 Nishi 8, Sapporo 060-8628, Japan.

3. Graduate School of Advanced Science and Engineering, Hiroshima University, Higashihiroshima, 1-3-1, Kagamiyama, Hiroshima, 739-8526, Japan

*Corresponding author: Y. Mizuguchi (mizugu@tmu.ac.jp)

**Abstract**

Compositionally complex alloys (CCAs) are a new category of high-entropy materials containing more than two high-entropy phases in an alloy sample. We applied the concept of CCA to a A15-type compound superconductor V$_3$X where X site is occupied by Al, Si, Ga, Ge, and Sn. The V$_3$X sample shows bulk superconductivity with a transition temperature ($T_c$) of 6.4 K, which is close to the $T_c$ of V$_3$Ge. The upper critical field ($B_{c2}(0)$) of V$_3$X estimated from magnetic susceptibility is 8.8 T, which is ~50% higher than that of V$_3$Ge ($B_{c2}(0)$ = 5.8 T). The presence of at least five different A15 phases with high configurational entropy of mixing were revealed. We consider that the examined V$_3$X sample is an example system of CCA-type superconducting intermetallic compound.

Inhomogeneity in superconductors have been extensively studied in the fields of condensed matter physics and superconductor application. From the perspectives of condensed matter physics, electronic inhomogeneity in high-transition-temperature (high-$T_c$) cuprates (particularly Bi-based systems) has been experimentally studied because it provides important information for understanding the mechanisms of superconductivity in the cuprates [1]. In addition, theoretical research has suggested the possibility of an enhancement of $T_c$ in superconductors with local inhomogeneity [2]. Furthermore, superconductors with lone-pair electrons sometimes exhibit anomalously disordered local structure, which has been typically observed in $BiS_2$-based layered superconductors in recent studies [3]. From the perspectives of superconductivity application, introduction of artificial pinning centers by addition of impurities, defects, or nanostructures has achieved a high critical current density ($J_c$) and a high upper critical field ($B_{c2}$) in various practical superconductors [4–7]. Hence, introduction and control of inhomogeneity of superconductors are important issues on both pure and applied sciences of superconductivity.

As a new material family of inhomogeneous superconductors, superconductive high-entropy alloys (HEAs) [8,9] have been extensively studied since the first report of superconductivity in $Ta_{34}Nb_{33}Hf_8Zr_{15}Ti_{11}$ in 2014 [10,11]. HEAs are typically defined as alloys composed of five or more different elements with a concentration range between 5 to 35 at%, and hence, HEAs have a high configurational entropy of mixing ($\Delta S_{mix}$) defined as $\Delta S_{mix} = -R \Sigma_i c_i \ln c_i$, where $c_i$ and $R$ are compositional ratio and the gas constant, respectively. Interestingly, the superconducting properties of HEAs are different from its bulk or thin films [11], suggesting obvious inhomogeneity on structural and/or electronic states in the HEAs. Furthermore, in our previous works, the concept of HEA has been applied to structurally more-complicated compound superconductors including high-$T_c$ cuprate $RE$123 [12], BiS$_2$-based layered system [13], NaCl-type chalcogenides [14-16], transition-metal zirconides [17,18], and A15-type Nb$_3X$ [19]. In those HEA-type compounds, various sensitivity of superconducting properties to an increase in $\Delta S_{mix}$ were revealed. Notably, a specific heat jump at $T_c$ becomes anomalously broad with increasing $\Delta S_{mix}$ in HEA-type compounds [19,20], which indicates the presence of highly inhomogeneous local structures (bonds) and anomalous electronic states. However, a clear enhancement of superconducting properties ($J_c$ or $B_{c2}$) by the effect of high entropy have not been observed in HEA-type superconductors, so far, although slight improvements of superconducting properties have been reported in some materials [16,21]. Therefore, the extension of the material category of inhomogeneous superconductors based on high configurational entropy of mixing is desired to establish new strategy for improving superconducting properties of practical-use superconducting materials and creating superconductors with novel superconducting states.

In this study, we focused on a compound superconductor having greater inhomogeneity than that of HEA-type superconductors studied so far. Recently, in the field of high-entropy materials, a new category of multi-phase high-entropy alloys has been proposed as *compositionally complexed alloys (CCAs)* [22]. According to Ref. 22, the CCAs have to possess an average composition similar to HEAs, which is more than 5 elements with less than 35 at%. In addition, multi-phase effects have to improve its functionality in CCAs. Here, we show the synthesis and the superconducting properties of an example of CCA-type compound superconductor with an A15-type structure. Being different from other HEA-type compound superconductors including Nb$_3X$ [23], the V$_3$Al$_{0.07}$Si$_{0.30}$Ga$_{0.08}$Ge$_{0.30}$Sn$_{0.25}$ (called V$_3X$ in this paper) sample shows inhomogeneous phase separation into at least five A15 phases with high $\Delta S_{mix}$ and different lattice parameters, which was detected by synchrotron X-ray diffraction (SXRD). Noticeably, the V$_3$X sample exhibits $B_{c2}(0)$ of 8.8 T, which is clearly higher than $B_{c2}(0)$ of ~7.3 T for V$_3$Ge with a similar $T_c$. With those experimental facts, we propose that the V$_3$X is an example of a CCA-type compound superconductor with a high $B_{c2}$ improved by the CCA effects.

The polycrystalline samples of V$_3$Al$_{0.07}$Si$_{0.30}$Ga$_{0.08}$Ge$_{0.30}$Sn$_{0.25}$ (V$_3$X) and V$_3$Ge were synthesized by arc melting. The raw materials were powders of V (99.9%, Kojundo Chemical Laboratory), Si (99.99%, Kojundo Chemical Laboratory), Ge (99.99%, Kojundo Chemical Laboratory), Sn (99.99%, Kojundo Chemical Laboratory), and Al (99.9%, Kojundo Chemical Laboratory) and grains of Ga (99.9999%, The Nilaco Corporation). The raw materials were pressed into a pellet and arc-melted in the Ar atmosphere. Melting was repeated two times after flipping. We compared the properties of the V$_3$X samples obtained by a mono-arc furnace and a tetra-arc furnace, which are labeled V$_3$X-A and V$_3$X-B, respectively. The temperature dependence of magnetic susceptibility was measured using a superconducting quantum interference devise (SQUID) (MPMS3, Quantum Design). The electrical resistivity was measured by four-probe method on PPMS (Quantum Design). The terminals were fabricated using an Ag paste and Au wires with a diameter of 25 μm. The specific heat under magnetic fields was measured by the relaxation method on PPMS. The chemical compositions of V$_3$X samples were evaluated by an energy-dispersive X-ray spectrometry (EDX; Oxford SwiftED on Hitachi-hightech TM3030). The SXRD patterns were obtained using the multiple microstrip detector MYTHEN system at the BL02B2 beamline of the synchrotron facility SPring-8 [24,25] with X-ray of 25 keV (wavelength, $\lambda$ = 0.495810(1) and 0.496118(1) Å) under proposal numbers 2021A1573 (for $V_3$Ge and sample A) and 2021B1175 (sample B). The phase split and the crystal structural parameters ware determined by the Le Bail method and the Rietveld method using the SXRD pattern in $2\theta < 74.2°$ ($d > 0.41$ Å). The Rietveld structure refinements were performed using the *JANA2020* software [26]. The images of the crystal structure were drawn using the *VEATA* software [27].

The EDX analysis revealed that the average compositions of $V_3X$-A and $V_3X$-B (normalized by V) are $V_3Al_{0.06}Si_{0.41}Ga_{0.07}Ge_{0.40}Sn_{0.16}$ and $V_3Al_{0.08}Si_{0.34}Ga_{0.03}Ge_{0.32}Sn_{0.19}$, respectively. The average compositions are close to the nominal value. The powder SXRD patterns of $V_3$Ge and $V_3X$ samples ($V_3X$-A and $V_3X$-B) are shown in Fig.1. The diffraction peaks of $V_3X$ were quite broad as compared to those of $V_3$Ge, and that was split into five peaks, which was confirmed by the Le Bail analysis. In the diffraction pattern for $V_3X$-B, main peaks were sharper than those for $V_3X$-A, but an impurity phase (VAl) was detected for the $V_3X$-B sample. Therefore, we performed the Rietveld refinement for $V_3X$-A to characterize the components causing the broad XRD peaks in $V_3X$. Figure 2 shows the Rietveld profile (fitting result) for $V_3X$-A and the schematic images of the structure models used for the analysis. An impurity phase of Sn with a population of 1.1% was detected. The broad peaks for $V_3X$-A were fitted by assuming five phases having different lattice parameters and compositional ratio of the $X$-site atom. The refinement results for five phases are summarized in Tables I, and the structural parameters used for the analysis are summarized in table II. In this analysis, the occupancy of Al and Ga were fixed as 0.07 and 0.08 (nominal values), respectively, because it is generally difficult to distinguish Si to Al and Ge to Ga by XRD because of similar atomic numbers. With increasing lattice parameter, the occupancy of Si and Al decreased, and that of Sn increased. The occupancy of Ge and Ga was almost constant for all the phases. The difference in lattice parameter of the five phases was caused by the average ionic radius of the contained elements at the $X$ site. As an important conclusion of the refinement, we revealed that (at least) five A15 phases with different lattice parameter and composition are included in the V$_3X$-A sample, which satisfies the structural condition of CCA.

To examine the difference in $B_{c2}$ between V$_3$Ge and V$_3X$, we measured magnetic susceptibility and electrical resistivity because improvement of functionality is required for CCA [22]. The temperature dependences of magnetic susceptibility under $B = 0$ T for V$_3$Ge, V$_3X$-A, and V$_3X$-B are displayed in Fig. 3(a). Superconducting transitions were observed at $T = 6.6$, 6.6, and 6.2 K for V$_3$Ge, V$_3X$-A, and V$_3X$-B samples, respectively. Although V$_3X$-A and V$_3X$-B exhibit similar $T_c$, the amplitude of the diamagnetic signals for V$_3X$-B is larger than that for V$_3X$-A. Since the estimated shielding volume fraction for V$_3$Ge and V$_3X$-A is close to 100%, superconductivity in all the samples are bulk in nature, and the large signal for V$_3X$-B would be caused by the demagnetization effect and/or the porosity of the sample. Figures 3(b)–3(d) show enlarged views under $B = 7$ T. Although the superconducting transition of V$_3$Ge is fully suppressed under 7 T at the temperature range of $T > 1.8$ K, the superconducting transitions are still observable for V$_3X$-A, and V$_3X$-B, indicating the improved $B_{c2}$ in the V$_3X$ samples as compared to V$_3$Ge. Figure 3(e) shows the magnetic field-temperature diagrams for V$_3$Ge and V$_3X$ samples. The $B_{c2}(0)$ are estimated as 5.8, 8.7, and 8.8 T for V$_3$Ge, V$_3X$-A, and V$_3X$-B, respectively. From the susceptibility measurements, we confirmed that $B_{c2}(0)$ for V$_3X$ is 50% higher than that for V$_3$Ge.

Figures 4(a)–4(c) show the temperature dependences of the electrical resistivity ($\rho$) under various magnetic fields for V$_3$Ge and V$_3X$ samples. At low temperatures near the superconducting transition, the normal-state $\rho$ for the V$_3X$ samples are clearly higher

than that for V$_3$Ge. A higher $\rho$ in HEA-type superconductors than that in zero- or low-entropy superconductors with the same type of material has been commonly observed in various HEA-type superconductors [23]. Under $B$ = 0 T, $\rho$ begins to decrease at $T_c^{onset}$ = 6.3, 6.4, and 6.5 K, for V$_3$Ge, V$_3X$-A, and V$_3X$-B samples, respectively. The $T_c$ decreases with an increase in magnetic field. Figure 4(d) shows the magnetic field-temperature diagram based on the magnetoresistivity data. The $B_{c2}$ (0) was estimated as 7.3, 8.8, and 8.9 T for V$_3$Ge, V$_3X$-A, and V$_3X$-B, respectively, which exhibits a ~20% higher $B_{c2}(0)$ for V$_3X$ than that for V$_3$Ge. The value of V$_3$Ge is comparable to the value reported in earlier works [28]. Although the improvement of $B_{c2}(0)$ determined from magnetoresistivity is smaller than that determined from susceptibility data, the trend of the improvement of $B_{c2}$ by the increased entropy was clearly confirmed in the V$_3X$ system. Here, the measurements of magnetic susceptibility and electrical resistivity confirmed that V$_3X$ samples exhibit an improved $B_{c2}(0)$ with respect to pure (zero-entropy) V$_3$Ge. Therefore, the examined V$_3X$ samples possess structural and superconducting properties enough for being regarded as a CCA-type compound superconductor. To the best of our knowledge, this is the first report on a CCA-type compound superconductor with detailed characterization on structure, including refinement with different phases, and superconducting properties.

To confirm the bulk nature of superconductivity in V$_3X$, we performed specific heat measurements. The temperature dependences of electronic specific heat capacity for superconducting states ($C_{SC}$) for V$_3$Ge, V$_3X$-A, and V$_3X$-B are displayed in Fig.4. The data of the total specific heat ($C$) and the electronic specific heat ($C_{el}$) for those samples are shown in Fig. S1 (supplemental data). and The $C_{SC}$ was calculated by subtracting the phonon contributions ($\beta T^3$) and normal-state electronic specific heat ($\gamma T$) from the total specific heat $C$, where $\gamma$ and $\beta$ are electronic specific heat parameter and that for phonon contributions: $\gamma$ = 32.33(7), 24.7(2), and 26.97(2) mJ / (mol K$^2$) and $\beta$ = 0.0764(8), 0.098(2), and 0.0917(3) mJ / (mol K$^4$) for V$_3$Ge, V$_3$X-A, and V$_3$X-B, respectively. The specific heat jump below $T_c$ for V$_3$Ge shows a sharp peak, but those for the V$_3$X samples have largely broadened structure. However, the results indicate that the entropy is clearly changing below $T_c$ for the V$_3$X samples. Similar trend on the broadening of the peak structure of $C_{SC}/T$ were observed in other HEA-type superconductors including Nb$_3$(Al$_{0.2}$Sn$_{0.2}$Ge$_{0.2}$Ga$_{0.2}$Si$_{0.2}$), Nb$_3$(Al$_{0.3}$Sn$_{0.3}$Ge$_{0.2}$Ga$_{0.1}$Si$_{0.1}$), and Co$_{0.2}$Ni$_{0.1}$Cu$_{0.1}$Rh$_{0.3}$Ir$_{0.3}$Zr$_2$ [17,19,20]. Since those HEA-type superconductors were characterized as single phases based on Rietveld refinements, the origin of the broad peaks of $C_{SC}$ observed in the present V$_3$X samples would be explained by both the effect of high entropy in the phases and the presence of phase separation. On the basis of the presence of entropy changes in specific heat data and large shielding fraction observed in susceptibility data, we consider that superconductivity in the examined V$_3$X samples is bulk in nature.

Herein, we discuss the effects of the mixing entropy and the presence of phase separation on the improved $B_{c2}$ in V$_3$X. The $B_{c2}$ of A15 superconductors is generally affected by the effect of the presence of martensitic transformation by the difference in the compositional ratio from stoichiometric composition and the addition of third element [29]. Both approaches also bring the disorder in the crystal structure of the A15 superconductor, due to the suppression of the martensitic transformation. The suppression of the structural phase transition temperature by the solid solution effect has been reported in various systems [30,31]. In the case of the present V$_3$X samples, both complex chemical disorder and phase separation are there. Hence, we consider that the high entropy effect, which results in high-entropy solution and the suppression of the phase transition, would be one of the origins of the improvement of $B_{c2}$ in the samples. Furthermore, although we have not characterized the domain sizes of the five different V$_3$X phases revealed by the refinement of the SXRD pattern, the optimization of the domain size is expected to further control the phase transition as studied in nanocrystalline materials [32,33]. On the basis of the discussion above, we propose that the improved $B_{c2}$ in the present V$_3$X samples can be ascribed to the fact that the V$_3$X samples are CCA type; the sample composition was initially designed based on the HEA strategy, phase separations were observed, and its functionality was improved. The application of the CCA concept to practical superconducting materials will be useful for improvement of the properties and will provide us with new pathway to develop new superconducting materials for application and superconductors with novel superconducting states.

In conclusion, we synthesized new HEA-type superconductor V$_3$X (V$_3$Al$_{0.07}$Si$_{0.30}$Ga$_{0.08}$Ge$_{0.30}$Sn$_{0.25}$) by arc melting and demonstrated the phase separation into five high-entropy A15 phases with different compositions of the X site in V$_3$X and the improvement of $B_{c2}(0)$ as compared to zero-entropy V$_3$Ge with a comparable $T_c$. We consider that the complex structure of V$_3$X in the CCA state mainly caused the suppression of the structural phase transition and improved $B_{c2}(0)$. The current investigation on the CCA-type A15 superconductor V$_3$X proposes that the introduction of CCA states into practical-use superconducting materials will be a new pathway for pure and applied superconductivity researches.

**Acknowledgements**

The authors thank O. Miura, Md. R. Kasem, and T. D. Matsuda for their support in experiments.

## Author contributions

Conception and experimental design: Y.N., A.Y., and Y.M. Carrying out measurements: all the authors. Manuscript composition: Y.N. and Y.M.

## Conflicts of interest or competing interests

The authors declare no competing interests.

## Data availability

The data presented in this paper can be accessed by contact to the corresponding author.

## Fundings

This work was partly supported by Grant-in-Aid for Scientific Research (KAKENHI) [Nos.: 18KK0076, 21H00151, 21K18834]; and Tokyo Metropolitan Government Advanced Research [H31-1].

**Tables**

Table I. Refinement results for the A15 phases refined for the SXRD data of V$_3$X-A. Space group $Pm\bar{3}n$; $Z = 2$; $wRp = 0.047$; $R_B = 0.066$-$0.107$; $R_F = 0.041$-$0.070$; goodness-of-fit = 5.52. In the analysis, the occupancies of Al and Ga of the X-site were fixed as 0.07 and 0.08, respectively.

| | $a$ (Å) | Volume ratio of phase | occupancy of Si and Al | occupancy of Ge and Ga | occupancy of Sn |
|---|---|---|---|---|---|
| Phase 1 | 4.77197(2) | 0.102(2) | 0.625(6) | 0.248(6) | 0.127(6) |
| Phase 2 | 4.78576(2) | 0.2285(7) | 0.538(4) | 0.359(4) | 0.103(4) |
| Phase 3 | 4.80356(3) | 0.1762(8) | 0.439(6) | 0.373(6) | 0.188(6) |
| Phase 4 | 4.82373(4) | 0.2046(9) | 0.473(6) | 0.311(6) | 0.216(6) |
| Phase 5 | 4.85697(3) | 0.2772(6) | 0.256(3) | 0.378(3) | 0.366(3) |

Table II. Structural parameters for V$_3$X. Atomic positions and isotropic thermal parameter ($U_{iso}$) were restricted as the same for all the A15 phases.

| Atom | Site | Symmetry | $x$ | $y$ | $z$ | $U_{iso}$ (Å$^2$) |
|---|---|---|---|---|---|---|
| V | 6c | $\bar{4}m2$ | 1/4 | 0 | 1/2 | 0.00860(6) |
| Si | 2a | $m\bar{3}$ | 0 | 0 | 0 | 0.00611(8) |
| Al | 2a | $m\bar{3}$ | 0 | 0 | 0 | = $U_{iso}$ (Si) |
| Ge | 2a | $m\bar{3}$ | 0 | 0 | 0 | = $U_{iso}$ (Si) |
| Ga | 2a | $m\bar{3}$ | 0 | 0 | 0 | = $U_{iso}$ (Si) |
| Sn | 2a | $m\bar{3}$ | 0 | 0 | 0 | = $U_{iso}$ (Si) |

**Figures**

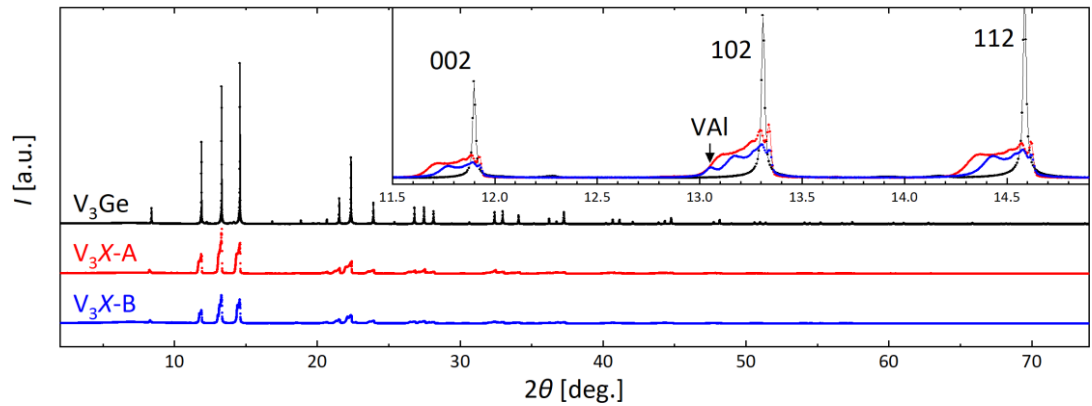

Fig. 1. Synchrotron powder X-ray diffraction (SXRD) patterns of V$_3$Ge, V$_3$X-A, and V$_3$X-B. The inset shows zoom on the (0 0 2), (1 0 2), and (1 1 2) peaks. For V$_3$X-B, a VAl impurity was detected as shown in the inset.

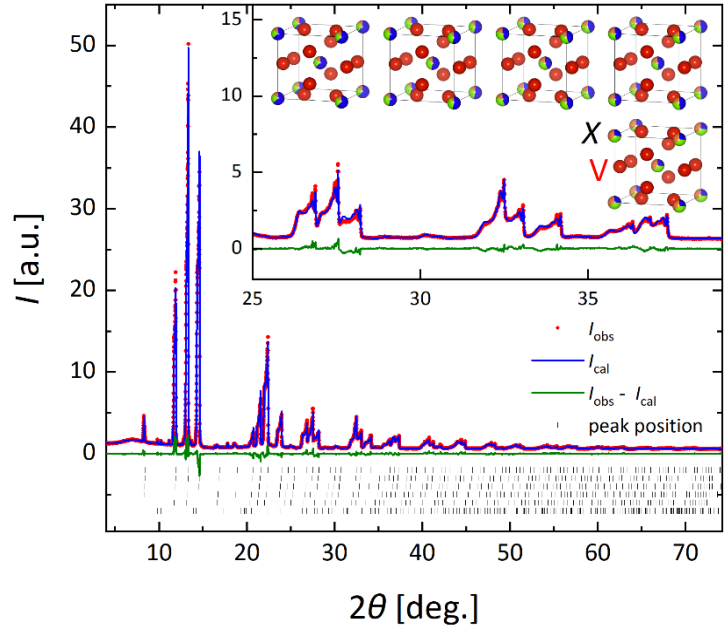

Fig. 2. Rietveld profile fitting result for V$_3$X-A and structure model of five phases obtained from the refinement. The red dots show the measured data, and the blue line is the fitted profile. The green line is the difference curve, and the black ticks are the peak positions of phase 1–5 and the impurity of Sn. The positions of the ticks for phase 1, 2, 3, 4, 5, and Sn are from top to bottom in the figure. The color of the X site in the model images shows the occupancy of atoms in X-site. The bule area corresponds to Si and Al, and the yellow-green area shows the occupancy of Ge and Ga. The orange area corresponds to the occupancy of Sn.

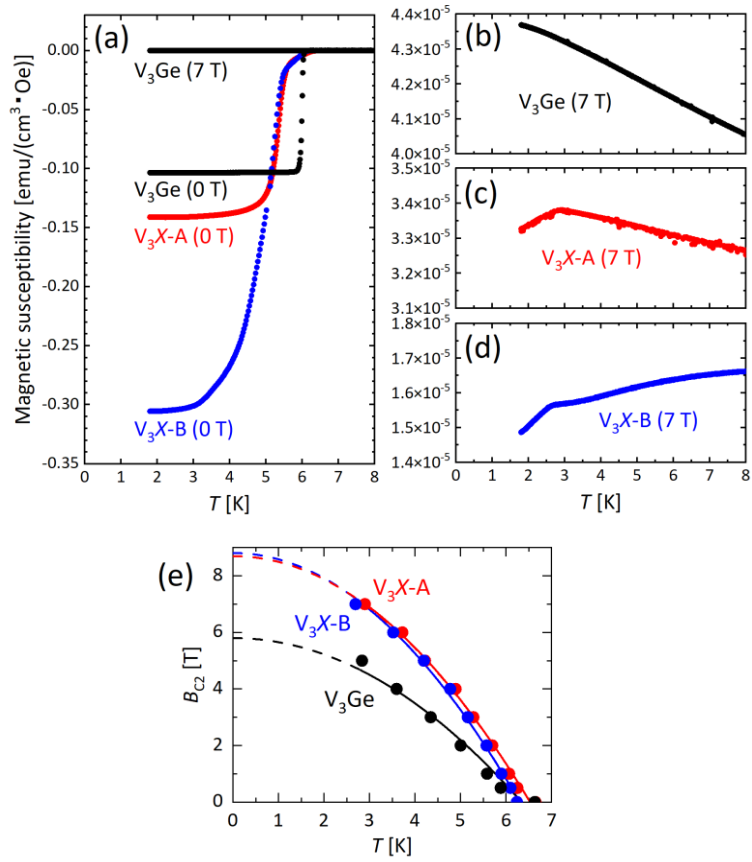

Fig. 3. (a) Temperature dependences of magnetic susceptibility for $V_3Ge$, $V_3X$-A, and $V_3X$-B under $B = 0$ T. (b–d) The temperature dependences of susceptibility under $B = 7$ T. (e) $B_{c2}-T$ plots for $V_3Ge$, $V_3X$-A, and $V_3X$-B.

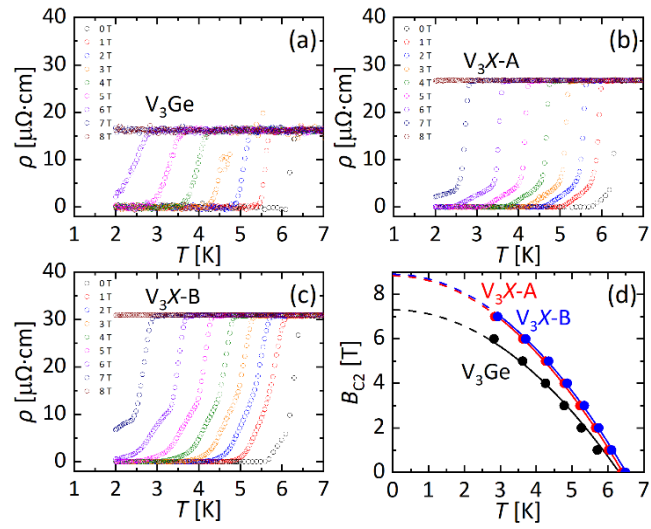

Fig. 4. Temperature dependences of electrical resistivity for (a) V$_3$Ge, (b) V$_3$X-A, and (c) V$_3$X-B. (d) $B_{c2}$-$T$ plots for V$_3$Ge, V$_3$X-A, and V$_3$X-B.

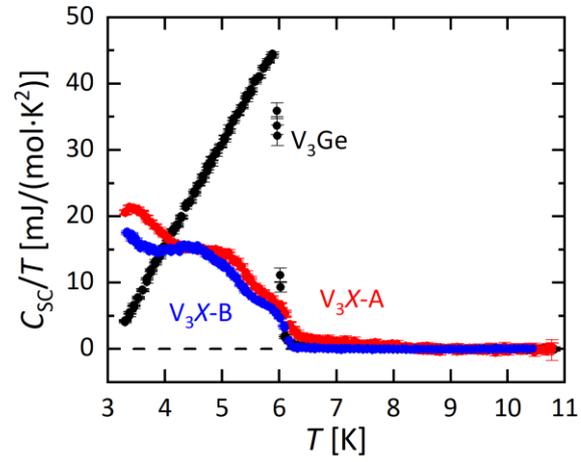

Fig. 5. Temperature dependences of electronic specific heat for superconducting states ($C_{SC}/T$) calculated from $C_{SC} = C - \gamma T - \beta T^3$ for V$_3$Ge, V$_3$X-A, and V$_3$X-B. The black, red, and blue circles are data for V$_3$Ge, V$_3$X-A, and V$_3$X-B, respectively.

# Supplemental data

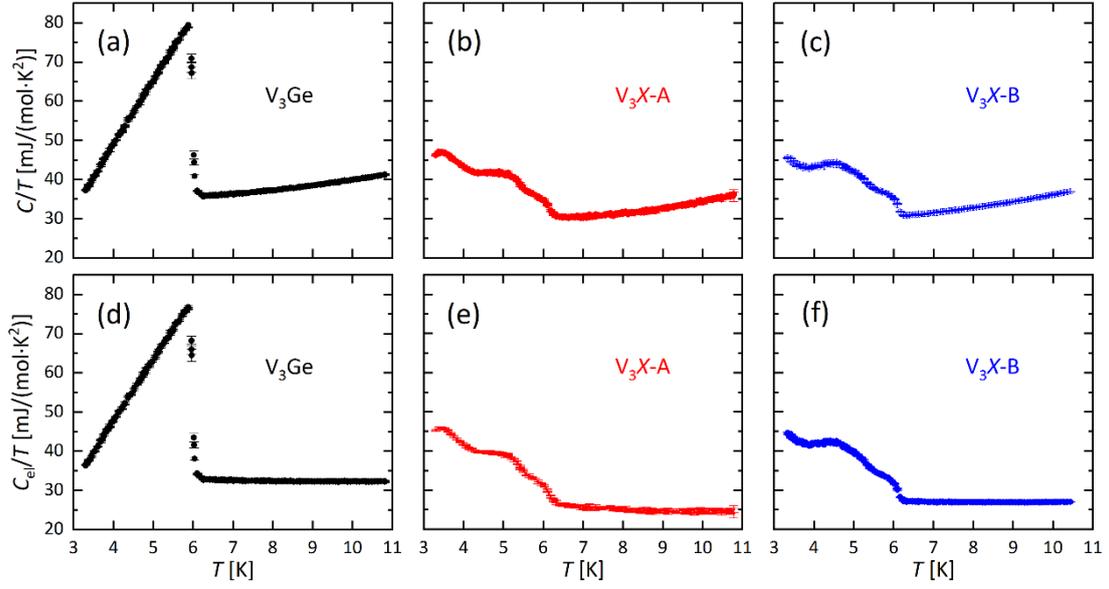

Fig. S1. Temperature dependences of (a–c) total specific heat ($C/T$) and (d–f) electronic specific heat ($C_{el}/T$), which was calculated from $C_{el} = C - \beta T^3$, for $V_3$Ge, $V_3X$-A, and $V_3X$-B. The black, red, and blue circles are data for $V_3$Ge, $V_3X$-A, and $V_3X$-B, respectively.